\documentclass{jaa}
\usepackage{natbib}
\usepackage{graphicx}

\begin{document}\sloppy

%%paper title
%%For line breaks \\ can be used within title
\title{Open cluster BSS dynamical clock dependence on the Milly Way gravitational field}

%%author names are separated by comma (,)
%%use \and before the last author name
%%use a * along with the number separated by comma
%% for the  author for correspondence
%%\textsuperscript{number} is used for affiliation
%%\affilOne, \affilTwo etc., upto \affilTwentyfive is possible
%%Please note the first letter after \affil is capitalised in the command
%%

\author{Andr\'es E. Piatti\textsuperscript{1,2,*}}
\affilOne{\textsuperscript{1}Instituto Interdisciplinario de Ciencias B\'asicas (ICB), 
CONICET-UNCUYO, Padre J. Contreras 1300, M5502JMA, Mendoza, Argentina; \\
\textsuperscript{2}Consejo Nacional de Investigaciones Cient\'{\i}ficas y T\'ecnicas 
(CONICET), Godoy Cruz 2290, C1425FQB,  Buenos Aires, Argentina}
%\affilTwo{\textsuperscript{2}Department of Q, University Z, Place Pincode, Country.}

%%escape two column mode for title, affiliation and abstract
%%by giving \twocolumn command as shown

\twocolumn[{

\maketitle

%%include \corres to print the corresponding author Email id
\corres{andres.piatti@fcen.uncu.edu.ar}

%%include \msinfo for
%%manuscript information such as
%%received, revised and accepted dates
%%
%\msinfo{1 January 2015}{1 January 2015}

%%abstract
\begin{abstract}
Since recent years, mass segregation driven by two-body relaxation in
star clusters has been proposed to be measured by the so-called dynamical clock, $A^+$, a
measure of the area enclosed between the cumulative radial distribution of blue straggler 
stars and that of a reference population. Since star clusters spend their lifetime
immersed in the gravitational potential of their host galaxy, they are also subject
to the effects of galactic tides. In this work, I show that the $A^+$ index of a star 
cluster depends on both, its internal dynamics as it were in isolation and on the effects
of galactic tides. Particularly, I focused on the largest sample of open clusters
harboring blue straggler stars with robust cluster membership. I found that these
open clusters exhibit an overall dispersion of the $A^+$ index in diagnostic diagrams where
Milky Way globular clusters show a clear linear trend. However, as also experienced
by globular clusters, $A^+$ values of open clusters show some dependence on their
galactocentric distances, in the sense that clusters located closer or farther that
$\sim$ 11 kpc from the Galactic center have larger and smaller $A^+$ values, respectively.
This different response to two-body relaxation and galactic tides in globular
and open clusters, which happen concurrently, can be due to their different masses.
More massive clusters can somehow protect their innermost regions from galactic tides
more effectively.
\end{abstract}

%%insert keywords separated by 3 hyphens using \keywords{words}

}]
%%close the twocolumn escape here

%%include \doinum{number}for the DOI number in the header
%%include \volnum{number} for the volume number in the header
%%include \year{yyyy} for  year of publication in the header
%%include \pgrange{num--num} page range of article in the header
%%include \artcitid{num} for the article citation id
%%include \lp to print last page of the article
%%include \setcounter{page}{pagenum} for the exact starting page of the article

%\doinum{12.3456/s78910-011-012-3}
%\artcitid{\#\#\#\#}
%\volnum{000}
%\year{0000}
%\pgrange{1--}
%\setcounter{page}{1}
%\lp{1}

\def\msun{\hbox{M$_\odot$}}
\def\prepare{\hbox{in preparation}}

\section{Introduction}

It is widely accepted that mass segregation is a phenomenon observed in collisional 
stellar systems as one manifestation of the internal two-body relaxation process. It consists
in kinematic energy exchanges between stars, which leads more massive stars to
slow down and to spiral toward the innermost regions, while less massive stars
increase their velocities reaching the outermost cluster regions \citep{alessandrinietal2014}.
From a theoretical point of view, the level of mass segregation in a star cluster is
given by the central relaxation time ($t_{rc}$), which depends on the core radius, the central
mass density, the average mass of the cluster stars, and the number of
cluster members \citep{djorgovski1993}. In practice, the $N_{relax}$ index is used instead,
defined as the ratio between the star cluster age and $t_{rc}$. Because of the wide
range of values of the measured parameters to compute $N_{relax}$,
different stellar systems show different levels of mass segregation.

In a star cluster, blue straggler stars (BSSs) are in general more massive stars than 
Main Sequence (MS) ones; hence they have extensively been used to measure the mass 
segregation level of stellar systems. They occupy a clearly identifiable region in the color-magnitude
diagram at brighter magnitudes and bluer colors than the position of the MS turnoff 
\citep[see, e.g.][]{lietal2023}. \citet{ferraroetal2018} measured $A^+$ -- the area 
enclosed between the cumulative radial distributions of BSSs and that of a reference (MS) population 
\citep[see][]{alessandrinietal2016} -- in  27 Milky Way old globular clusters, and found that $A^+$
correlates with $N_{relax}$. From then on, $A^+$ has been referred to as a star cluster 
dynamical clock. It has been used to measure the dynamic evolutionary stage of open clusters
\citep[e.g.,][]{raoetal2021,raoetal2023a}. Indeed, different studies on BSSs in Milky Way open 
clusters have recently been published, with several updated compilations and analysis of 
BSS properties \citep{js2021,raoetal2023,rainetal2024}. 

Because a star cluster moves in an external gravitational potential, tides also
affect the spatial distribution of its stars and therefore the cluster structural parameters. 
\citet{piattietal2019b} showed that the core, half-mass and Jacobi radii of the
Milky Way globular clusters are in general larger as their galactocentric distances increase,
because at larger galactocentric distances the Milky Way potential weakens,
allowing the cluster to expand. Furthermore, the pace of change with the galactocentric distance is 
more important for Jacobi radii than for core ones. \citet{pm2018} also studied the
correlation between the cluster size and its position in its host galaxy by analyzing the population
of Large Magellanic Cloud globular clusters, and found similar results. The cluster radius change 
caused by the host galaxy gravitational field implies that the cluster is dynamically accelerated,
i.e., it is in a more advanced  internal dynamics evolutionary stage in the presence of 
stronger tides. Consequently, tides from an external gravitational field affect the level of mass 
segregation in a star cluster \citep[see also][]{piatti2020c}.

\begin{figure*}
\includegraphics[width=\textwidth]{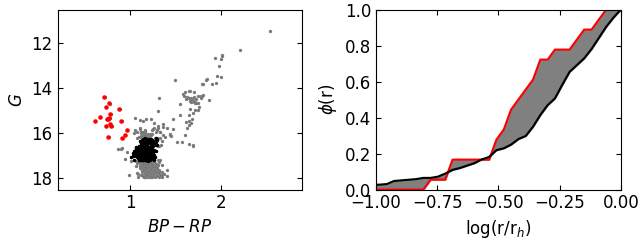}
\caption{{\bf Left panel:} {\it Gaia} color-magnitude diagram of NGC~2158 (gray points)
for stars located inside $r50$. Red and black 
points represent cluster BSSs and reference population stars with $P$ $>$ 70$\%$, respectively,
located inside $r50$. {\bf Right panel:}
cumulative spatial distribution function of BSSs and reference population stars drawn with
red and black lines, respectively. The gray area enclosed by them represents the $A^+$ index.}
\label{fig1}
\end{figure*}

Open clusters are less massive than globular clusters, so that mass segregation is 
expected to be more affected by galactic tides. This would imply that the $A^+$ index needs to be 
corrected by tidal effects previously to be used as a cluster dynamical clock. Precisely, 
we analyze this phenomenon in this work. In Section~2 we describe the updated homogeneous compilation  
of BSSs in open clusters and the 
measurement of $A^+$ indices for robust sample of open cluster. From them, 
in Section~3, we analyze the dependence of $A^+$
with the position of the open clusters in the Milky Way, and discuss the outcomes.
Section~4 summarizes the main conclusions of this work.

\section{The $A^+$ index}

\citet[][see their figure 4]{raoetal2023a} analyzed 21 open clusters and showed that
the correlation between $A^+$ and $N_{relax}$ is moderate (0.5), with a relative
wide range of internal dynamics evolutionary stages at any $A^+$ values,
for 0.0 $\le$ $A^+$ $\le$ 0.2. They arrived at this result using {\it Gaia} EDR3
data \citep{gaiaetal2021} to estimate cluster properties (half-light radius, 
central mass density, average stellar mass, etc) from identified members.
In order to probe whether such a dispersion has its origin in any dependence of the $A^+$ index 
on the effects of tides caused by the Milky Way gravitational potential, we decided
to use the updated catalog of populations of BSSs in open clusters compiled by
\citet{rainetal2021}. The catalog is the result of an homogeneous search for BSSs
using {\it Gaia} photometry, proper motions, and parallaxes, along with solid assessments
on their cluster membership. They used the membership 
probabilities obtained by \citet{cantatgaudinetal2020} from {\it Gaia} DR2 dat sets. 
The stringent criteria applied allowed \citet{rainetal2021}
to identify 897 BSSs in 408 open clusters. Because most of the clusters contain very few
number of BSSs, they concluded that open clusters are not a preferable environment for these 
kinds of stars.

From their catalog, we retrieved R.A. and Dec. coordinates, $G$ magnitude and $BP-RB$ color, and 
membership probability ($P$ ($\%$)) for every BSS with $P$ $>$ 70$\%$. We used this frequently
employed membership cut in star cluster photometric studies as a compromise between minimizing 
field contamination (e.g. $P$ $<$ 50$\%$) and maximizing the presence of highly ranked cluster 
members (e.g., $P$ $>$ 90$\%$). Nevertheless, in the subsequent analysis we compare the results
by using stars with $P$ $>$ 70$\%$ and $P$ $>$ 90$\%$. With the aim of assuring also a
reliable cluster statistics, we kept clusters with a number of BSSs inside their 
 half member radius \citep[$r50$;][]{cantatgaudinetal2020} larger than 5,  and
compiled a cluster sample as large as possible (see Table~\ref{tab1}). In the subsequent analysis, 
it is shown that this lower limit does not affect the results. We found a total 
of 18 open clusters that comply with such a requirement. We also retrieved  R.A. and Dec. 
coordinates, $G,BP,RP$  photometry, and membership probabilities $P$ for stars with 
$P$ $>$ 70$\%$ from the catalog compiled by \citet{cantatgaudinetal2020} of stars
belonging to a reference cluster stellar population for these 18 open clusters 
(see below). We note that both the data sets and the membership probabilities of the 
BBS and reference population stars come from the same source, namely {\it Gaia} DR2 and the
\citet{cantatgaudinetal2020}'s membership assessment. We then followed the recipe outlined 
by \citet{ferraroetal2018} to compute $A^+$, namely: i) to build the cumulative spatial distribution 
function of BSSs distributed within a circular area of radius equals to the cluster half
member
radius; ii) to construct the corresponding cumulative spatial distribution function of a reference 
cluster stellar population; and iii) to compute the value of the area enclosed between 
both cumulative spatial distribution functions.  For the calculation of the $A^+$ index and the
subsequent analysis, we adopted the cluster  central coordinates and cluster half member radii 
given by  \citet{cantatgaudinetal2020} (see Table~\ref{tab1}).

In practice, we first matched the \citet{cantatgaudinetal2020} and \citet{rainetal2021} catalogs,
and for each matched cluster we computed the distance to the cluster center of every star with
$P$ $>$ 70$\%$, and kept those located inside $r50$. Then, we sought for the faintest cluster BSS 
and used the corresponding $G$ magnitude to define the reference population, which consisted 
of cluster MS stars located inside $r50$ and with $G$ magnitudes down to 1 mag fainter than 
the faintest BSS $G$ mag. This criterion allows to maximize the use of the largest number
of BSSs, which can be unevenly distributed along the $G$ magnitude when comparing one cluster to another,
and thus more physically meaningfully to define the sample of reference population stars as those
stars less massive than the least massive BSS star. We found that using a $G$ mag interval of 0.5 mag
does not imply changes in the $A^+$ trend with respect to different cluster properties.
 We applied this criterion to define the reference population of the 18 
selected open clusters. We verified that the faintest MS stars used are brighter than $G$ $\sim$ 
18.0  mag. Figure~\ref{fig1} illustrates for NGC~2158 the placement of BSSs and MS stars belonging to the 
reference population, highlighted with red and black points, respectively. Similar plots for
the whole cluster sample are included in the Appendix.

\begin{figure}
\includegraphics[width=\columnwidth]{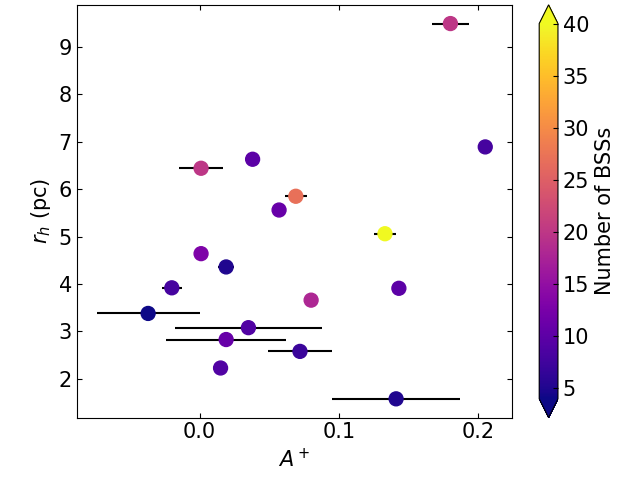}
\caption{Relationship between the $A^+$ index and $r50$ for the studied cluster sample.
Error bars are indicated.}
\label{fig2}
\end{figure}

We cumulatively added the stars using their distances to the cluster centers ($r$) as a variable and built the respective distribution function 
$\phi$($r$) in terms of log($r/r50$), where $r50$ is the half  member radius. The right 
panel of Figure~\ref{fig1} illustrates, as an example, the construction of the cumulative 
spatial distribution functions for BSSs and reference MS stars of NGC~2158  (see also the
cumulative functions for the whole cluster sample in the Appendix). As can be seen,
BSSs in NGC~2158 are more centrally concentrated than reference MS stars.  The 
area embraced by these curves represents $A^+$, which we obtained by calculating the 
difference between the integrals of both curves. By definition, the larger the $A^+$ value,
the more advanced level of cluster mass segregation.  Figure~\ref{fig2} shows that 
we equally employed relatively smaller and larger numbers of BSSs and $r50$ to compute $A^+$, 
which secures our analysis against any bias in the $A^+$ calculation procedure. The different
number of BSSs used (from 5 up to 39) is represented with filled circles of different sizes.
The uncertainties in 
the computed $A^+$ values were estimated assuming that the main source of  dispersion in 
the $A^+$  values is the number of BSSs and reference MS stars used. Therefore, we  repeated 
20 times the procedure of computing $A^+$, by considering numbers of stars whose membership
probabilities randomly vary from 70$\%$ up to 90$\%$, and adopted the largest difference between the
obtained $A^+$ value and that for $P$ $>$ 70$\%$ as the uncertainty in $A^+$ 
($\sigma$($A^+$)). Table~\ref{tab1} lists the resulting $A^+$ values with their associated 
errors and relevant cluster properties used in this work. A comparison of the present
$A^+$ values and those derived by \citet{raoetal2023a} for open clusters in common is shown in
Figure~\ref{figa18}. \citet{raoetal2023a} also computed $N_{relax}$ for their smaller
cluster sample. Computing $N_{relax}$ is not a straightforward task, because it depends
on the knowledge of the mean cluster mass, the central cluster stellar density, and the cluster core
radius, which can be reliably estimated from photometric data corrected for incompleteness.
Being this not the case for our entire cluster sample, nor that of \citet{raoetal2023a}
photometry (see also differences with \citet{raoetal2021}), we preferred to conduct
the subsequent analysis using widely used cluster structural parameters \citep{kharchenkoetal2013}.

%I found 18 open clusters in common with \citet{raoetal2023a}, who derived $A^+$ and $N_{relax}$ 
%values for them. However, their $A^+$ values are not directly comparable to ours, because they 
%used their own identification of BSSs and allowed stars with $P$ $>$ 0.2 to be cluster members. 
%Both reasons straightforwardly affect the calculation of the $A^*$ index, because of the different
%spatial distribution of different selected samples of BSSs in a star cluster. 
%Figure~\ref{fig3} shows the number of BSSs included in the catalog of \citep{rainetal2021} 
%compared to that used by \citet{raoetal2023a}. As can be seen, 
%there is a significant difference between both compilations of BSSs. Furthermore,
%11 of these 18 open clusters were previously analyzed by\citet{raoetal2021}, and 
%the number of BSSs used and derived $A^+$ values are notably different from those in
%\citet{raoetal2023a}. On the other hand, including stars with $P$ $>$ 20$\%$ implies
%a high probability to contaminate the cluster star samples with field stars.

\section{Analysis and discussion}

\begin{figure}
\includegraphics[width=\columnwidth]{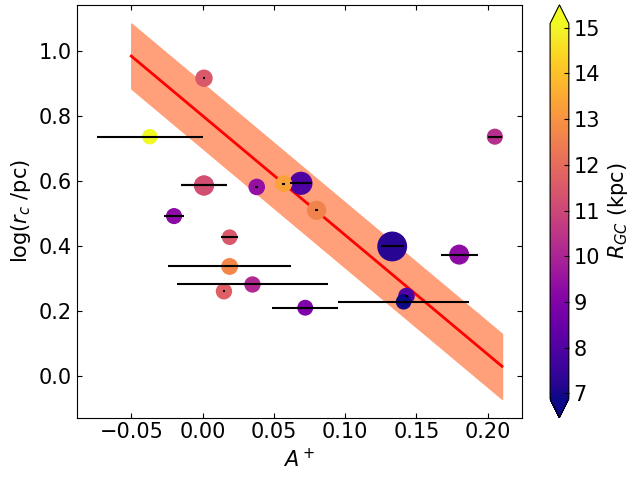}
\caption{Relationship between the $A^+$ index and $r_c$ for the studied cluster sample.
Error bars are indicated. The size of the circles is proportional to the number of
BSSs used to compute $A^+$. The solid line and shaded region represent the fitted relationship
and its uncertainty found by \citet{ferraroetal2018} for the Milky Way globular cluster
population. Pearson, Spearman, and p-value (T-test) are -0.12, -0.12, and 3$\times$10$^{-15}$,
respectively.}
\label{fig3}
\end{figure}

\begin{figure}
\includegraphics[width=\columnwidth]{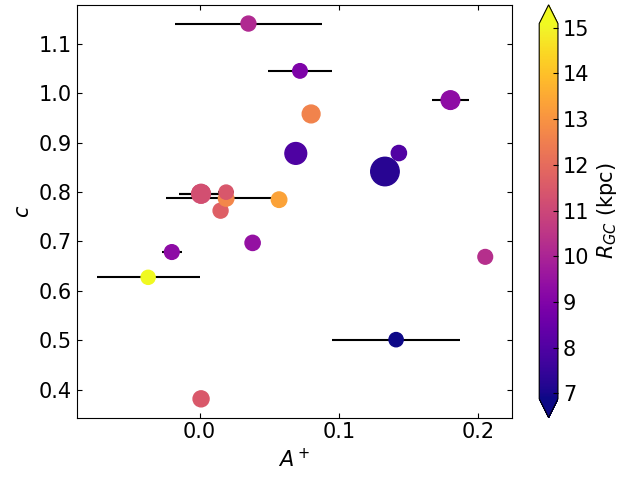}
\caption{Relationship between the $A^+$ index and the concentration parameter $c$ for 
the studied cluster sample. Error bars are indicated. The size of the circles is proportional 
to the number of BSSs used to compute $A^+$. Pearson, Spearman, and p-value (T-test) are 0.06, 0.11, 
and 9$\times$10$^{-133}$, respectively.} 
\label{fig4}
\end{figure}

We started our analysis by building Figure~\ref{fig3}, which depicts the correlation between
the core radius ($r_c$) derived by \citet{kharchenkoetal2013} and our resulting $A^+$ values 
for the 18 studied open clusters. Both quantities were obtained independently one to each other
and from homogeneous procedures. For the sake of the reader,
we draw the relationship between both parameters and its standard deviation derived
for Milky Way globular clusters by \citet{ferraroetal2018}. They also showed that
$N_{relax}$ correlates with $A^+$, which we do not examine in this work because only
some few open clusters have reliable $N_{relax}$ estimates. We should expect that $A^+$ values 
of open clusters follow the linear curve drawn in Figure~\ref{fig3}, if $A^+$ results in a 
indicator of the level of 
the cluster mass segregation driven by two-body relaxation as the cluster were in isolation. 
As can be seen, the studied open cluster sample shows
an important dispersion around the fitted relationship. With the aim of disentangling the
origin of such a dispersion, we colored the points according to the respective cluster
galactocentric distance. At first glance, farther open clusters tend to have smaller
$A^+$ values in comparison with those of open clusters closer to the Galactic center.
Figure~\ref{fig4} shows the behavior of the dynamical profile of a cluster given by the 
concentration parameter $c$ = log($r_t$/$r_c$), where $r_t$ is the tidal radius derived by 
\citet{kharchenkoetal2013}, from which is also seen the lack of correlation with
the $A^+$ index.

\begin{figure}
\includegraphics[width=\columnwidth]{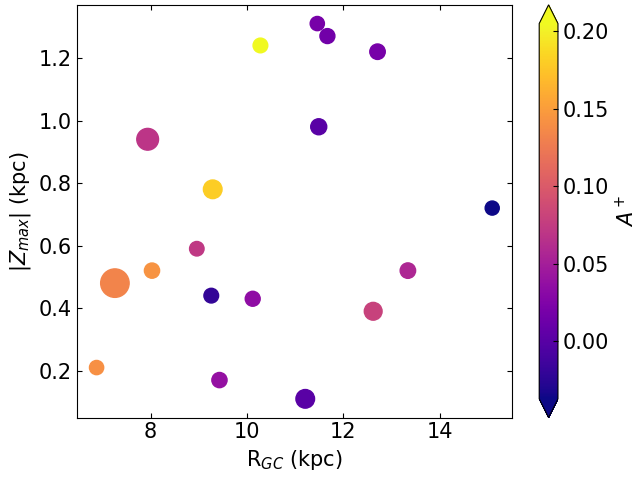}
\caption{Relationship between the galactocentric distance ($R_{GC}$) and the
maximum height above the Galactic plane ($Z_{max}$) for the studied cluster sample.
The size of the circles is proportional to the number of
BSSs used to compute $A^+$.}
\label{fig5}
\end{figure}

Based on Figures~\ref{fig3} and \ref{fig4}, we examined the correlation between the present 
$A^+$ values with the galactocentric distance ($R_{GC}$) and the maximum height above the
Galactic plane \citep[$Z_{max}$;][]{tarricgetal2021} for the studied cluster sample. Figure~\ref{fig5} shows the resulting 
relationship, which reveals that the $A^+$ index is prone to variation of the galactocentric 
distance, in the sense that the larger the galactocentric distance, the smaller the $A^+$ index.
In the case of the maximum height above the Galactic plane some trend is also glimpsed, although
it depends on $R_{GC}$, namely: for large $Z$ values (and large $R_{GC}$ ones) $A^+$ values
are generally smaller than those for smaller $Z$ values (and smaller $R_{GC}$ ones).
The observational trend supports the idea that the Milky Way galactic potential 
causes --through the effects of tides-- that open clusters closer to the Galactic center
have larger $A^+$ values in comparison with their counterparts
located farther from the Galactic center; mass segregation as measured by $r_c$ and
by the concentration parameter $c$ showing an overall dispersion. This result would seem to
suggest that the $A^+$ index  for open clusters is one of the primary drivers of the effects of galactic
tides rather than by $r_c$ and $c$, which according to \citet{ferraroetal2018} is sensitive to 
mass segregation in Milky Way globular clusters. As far as we are aware, our outcome is 
based on the largest sample of open clusters with $A^+$ index derived homogeneously. Particularly,
we emphasize on the solid assessments of BSS and reference MS cluster star membership
\citep{rainetal2021}.

Because $A^+$ values decrease with increasing galactocentric distances,
as also the effects of galactic tides do, we draw the conclusion  -- although from a
limited sample of 4 open clusters with 12 $<$ $R_{GC}$ (kpc) $<$ 15 -- that typical $A^+$
value of open clusters in the presence of a relative light external galactic field 
(or the absence thereof) is $\sim$ 0.0$\pm$0.05 (see Figure~\ref{fig5}).
This range ($\Delta$$A^+$ = 0.05) is nearly 10 times smaller than that found for Milky Way 
globular clusters, many of them populating the Galactic halo \citep{ferraroetal2018},  
which suggests that the $A^+$ index for the globular and open cluster populations
behaves differently. From this speculation, larger $A^+$ values in open clusters
would come from the effect of the galactic tides, thus mimicing the more advanced mass segregation 
level of an open cluster dynamically evolving in isolation. Note, however, that
realiable $N_{relax}$ values are needed in order to probe whether for a given $N_{relax}$ value, 
the $A^+$ index clearly decreases with $R_{GC}$. Nevertheless, if an attempt of a
linear relationship between $A^+$, $R_{GC}$, and $Z_{max}$ were suggested, as:

\begin{equation}
A^+ = a_1 + a_2 \times R_{GC} + a_3 \times Z 
\end{equation}

\noindent we would obtain $a_1$ = 0.244$\pm$0.069, $a_2$ = -0.019$\pm$0.006, and 
$a_3$= 0.030$\pm$0.038, with $\chi$$^2$ = 0.004.

We explored the different spatial distributions of BSSs and MS reference stars in open
clusters with smaller and larger $A^+$ values using  numerical simulations. We adopted
a King's \citep{king62} profile to describe the stellar density distribution and played with 
different combinations of $r_c$ and $r_t$ values. We
%counted the number of synthetic stars contained in different bins of each radial stellar density distribution function and 
built the cumulative distribution function as described in Section~2. The values of $r_c$ and $r_t$ 
used for the reference MS stars are those coming from the average of core and tidal radii
homogeneously derived for 2006 open clusters by \citet{kharchenkoetal2013}. We found that mean 
core and tidal radii increase from $\sim$ 0.8 pc up to 3.0 pc and from $\sim$ 9.0 up to 50.0 pc,
respectively, for $R_{GC}$ between 7 kpc and 16 kpc. As can be seen, open clusters
show a trend of increasing radii with the galactocentric distance, similarly to 
Large Magellanic Cloud and Milky Way globular clusters \citep{pm2018,piattietal2019b}. 
Furthermore, the $r_t/r_c$ ratio also increases with increasing galactocentric distances.

\begin{figure}
\includegraphics[width=\columnwidth]{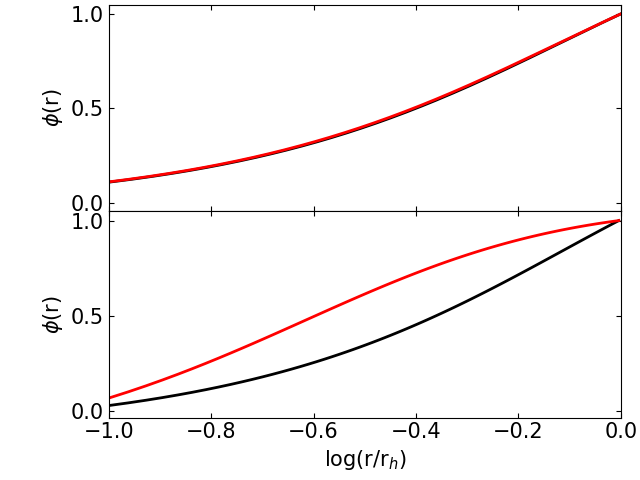}
\caption{Cumulative distribution functions of synthetic BSS and MS reference stars
drawn with red and black lines, respectively. Top and bottom panels correspond
to similar and different density distributions of BSS and MS reference stars,
respectively (see text for details).}
\label{fig6}
\end{figure}

Figure~\ref{fig5} shows that outer open clusters have $A^+$ $\sim$ 0.0. This implies that BSSs 
and MS reference stars are spatially distributed similarly, and hence both populations
can be modeled with the same $r_c$ and $r_t$ values corresponding to large galactocentric
distances. Figure~\ref{fig6} (top panel) shows the expected resulting cumulative distribution 
functions. In order to get $A^+$ $\sim$ 0.15 for inner galactic disk open clusters 
($R_{GC}$ = 7 kpc, see Figure~\ref{fig5}), we used the $r_c$ and $r_t$ values mentioned above 
for the MS reference stars, while for BSSs we used $r_c$ = 0.3 pc and ($r_t/r_c$)$_{BSS}$ = 
($r_t/r_c$)$_{MS}$= 3.75.
Figure~\ref{fig6} (bottom panel) depicts the two different cumulative distribution functions.
Here, $r_c$(BSS) plays an important role. We obtained similar $A^+$ values using $r_t$(BSS) as
large as $r_t$(MS). However, larger $r_c$(BSS) values lead to $A^+$ ones smaller than 0.15.
The above simulations show that the core radius of the MS reference stars is $\sim$ 2.7 times
larger than that of the BSSs to produce $A^+$ = 0.15. 

When dealing with Milky Way globular clusters, the difference between the core radii of the BSSs 
and MS reference stars, which implies mass segregation, is a function of the cluster core radius 
(see straight line in Figure~\ref{fig3}). The smaller the cluster core radius, 
the larger the level of mass segregation (larger $A^+$ values). At the same time, globular cluster 
core radii vary with the galactocentric distance as shown by \citep{piattietal2019b}, adding a 
residual behavior of $A^+$ as a function of the galactocentric distance 
\citep[see eq. (2) of][]{piatti2020c}. Here, Figure~\ref{fig3} shows that the $A^+$ index of
open clusters does not directly depends  on $r_c$. However,
open clusters can exhibit some level of mass segregation, i.e., different spatial distributions
of BSSs and MS reference stars, when comparing those located closer and farther than $\sim$ 11 kpc
from the Galactic center (see Figure~\ref{fig5}). This trend suggests that the galactic tides
are stronger drivers of mass segregation in open clusters than the two-body relaxation process.

\section{Conclusions}

It has been widely accepted that two-body relaxation led collisional systems
to experience mass segregation. Recently, the $A^+$ index was introduced with 
the aim of providing a direct measure of the level of mass segregation of
Milky Way globular clusters. Since then, $A^+$ was called a dynamical clock.
However, star clusters usually are found immersed in the gravitational
potential of their host galaxies, so that they are also subject of galactic
tide effects. One of the manifestations of the galactic gravitation forces 
acting on star clusters are the observed tidal tails. 

Based on this evidence, an avoidable question arises: at what level the
star cluster stellar density distribution is affected by galactic tides?
In order to contribute to its answer, we embarked in the analysis of
Milky Way and Large Magellanic Cloud globular and open clusters by examining
the dependence of the $A^+$ index on the galactocentric distance.
We found that $A^+$ values of globular clusters depend on both the level
of two-body relaxation and their position in the galaxy 
\citep{piatti2020c}. In this present work, we analyzed the behavior of the 
$A^+$ index for a sample of Milky Way open clusters. 

We used a recent updated catalog of BSSs in open clusters \citep{rainetal2021}
with robust assessments on their cluster membership. The final list of selected
clusters contains 18 objects with more than 5 BSSs with membership probabilties
$>$ 70$\%$ and {\it Gaia} photometry; same requirements were fulfilled for
cluster MS stars used as a reference stellar population. The derived $A^+$ 
values lead to conclude that open clusters do not show the trend with $r_c$
shown for globular clusters, but an overall dispersion. Nevertheless,
more open clusters are needed to be analyzed in order to make more definitive 
statements on this behavior. However, the residual correlation of the $A^+$
index with the cluster position in the galaxy seen for globular clusters
is also apparent in the studied open cluster sample. We think that the
different observed response of star clusters' internal dynamics to two-body 
relaxation and to galactic tides is related to their masses. More massive
clusters (globular versus open clusters) can somehow protect their innermost 
regions from galactic tides more effectively.

\begin{table*}
\caption{$A^+$ estimates and astrophysical parameters for the studied cluster sample. $N_{BSS}$ and
$N_{ref}$ refer to the number of BBS and reference stars with $P$ $>$ 70$\%$ located inside $r50$, respectively.}
\label{tab1}
\begin{small}
\begin{tabular}{lrrccccrrccc}\hline\hline
Cluster & R.A. & Dec. & log($t$ /yr) & $r_c$ & $r50$ & $r_t$ & $R_{GC}$ & Z & $A^+$ & $N_{BSS}$ & $N_{ref}$\\
        & (deg)& (deg)&              & (pc)  & (pc)  & (pc)  & (kpc)    & (kpc) &  &    &\\\hline

%Berkeley~12   &    71.100 &  42.691 & 9.60 & 1.96 & 2.07 & 18.00 & 12.49 & -0.15 & 0.014$\pm$0.014 & 8 & 32 \\
%Berkeley~14   &    74.935 &  43.488 & 9.20 & 3.24 & 4.86 & 25.72 & 13.13 &  0.06 & 0.105$\pm$0.050 & 7 & 117\\
Berkeley~17   &    80.130 &  30.574 &10.00 & 1.82 & 2.23 & 10.53 & 11.67 & -0.21 & 0.015$\pm$0.005 & 9 & 81\\
%Berkeley~18   &    80.531 &  45.442 & 9.63 & 9.21 &11.84 & 44.87 & 13.81 &  0.49 & 0.031$\pm$0.021 & 20& 71\\
%Berkeley~19   &    81.014 &  29.575 & 9.40 & 1.32 & 3.10 & 13.84 & 14.89 & -0.41 & 0.041$\pm$0.010 & 6 & 13\\
Berkeley~31   &   104.406 &   8.285 & 9.31 & 5.44 & 3.38 & 23.07 & 15.09 &  0.64 &-0.037$\pm$0.037 & 5 & 18\\
%Berkeley~32   &   104.530 &   6.433 & 9.70 & 2.71 & 5.49 & 18.28 & 11.14 &  0.24 & 0.159$\pm$0.006 & 23& 149\\
Berkeley~39   &   116.702 &  -4.665 & 9.90 & 8.23 & 4.64 & 19.81 & 11.49 &  0.69 & 0.001$\pm$0.004 & 13 & 123\\
Berkeley~99   &   350.260 &  71.778 & 9.50 & 2.67 & 4.36 & 16.84 & 11.46 &  0.90 & 0.019$\pm$0.006 & 5 & 19 \\
%Collinder~74  &    87.170 &   7.374 & 9.11 & 2.22 & 2.90 & 10.44 & 10.69 & -0.45 & 0.078$\pm$0.010 & 5& 74  \\
Collinder~261 &   189.519 & -68.377 & 9.95 & 2.50 & 5.06 & 17.36 &  7.26 & -0.28 & 0.133$\pm$0.010 & 40  & 467 \\
King~11	      &   356.912 &  68.636 & 9.04 & 1.91 & 3.08 & 26.43 & 10.12 &  0.35 & 0.035$\pm$0.053 & 9 & 76\\
Melotte~66    &   111.573 & -47.685 & 9.53 & 5.44 & 6.89 & 25.38 & 10.28 & -1.19 & 0.205$\pm$0.005 & 8 & 227\\
%Melotte~71    &   114.383 & -12.065 & 8.37 & 2.12 & 3.24 & 20.92 &  9.87 &  0.17 & 0.155$\pm$0.002 & 5&  30 \\
NGC~188       &    11.798 &  85.244 & 9.88 & 2.36 & 9.49 & 22.86 &  9.28 &  0.65 & 0.180$\pm$0.014 & 20 & 222\\
NGC~1193      &    46.486 &  44.383 & 9.70 & 2.17 & 2.83 & 13.29 & 12.70 & -1.05 & 0.019$\pm$0.043 & 11 & 60 \\
%NGC~1245      &    48.691 &  47.235 & 9.03 & 2.73 & 5.71 & 30.93 & 11.12 & -0.50 & 0.130$\pm$0.040 & 5& 27\\
%NGC~1798      &    77.914 &  47.691 & 9.25 & 3.24 & 3.57 & 22.06 & 13.27 &  0.43 & 0.009$\pm$0.047 & 22& 63\\
NGC~2141      &    90.734 &  10.451 & 9.23 & 3.90 & 5.56 & 23.75 & 13.34 & -0.52 & 0.057$\pm$0.003 & 11 & 156\\
NGC~2158      &    91.862 &  24.099 & 9.02 & 3.23 & 3.66 & 29.33 & 12.62 &  0.13 & 0.080$\pm$0.002 & 18 & 298\\
%NGC~2204      &    93.882 & -18.670 & 8.89 & 3.55 & 6.48 & 17.20 & 11.34 & -1.11 & 0.142$\pm$0.002 & 7& 194\\
%NGC~2243      &    97.395 & -31.282 & 9.03 & 1.64 & 3.57 & 20.99 & 10.58 & -1.15 &-0.018$\pm$0.025 & 14& 208\\
%NGC~2354      &   108.503 & -25.724 & 9.12 & 9.13 &   -0.058$\pm$0.010\\
%NGC~2477      &   118.046 & -38.537 & 9.00 & 8.85 &   -0.008$\pm$0.050\\
%NGC~2506      &   120.010 & -10.773 & 9.00 & 3.61 & 5.42 & 25.25 & 10.62 &  0.55 & 0.123$\pm$0.001 & 14& 396 \\
NGC~2682      &   132.846 &  11.814 & 9.45 & 1.62 & 2.58 & 17.98 &  8.96 &  0.47 & 0.072$\pm$0.023 & 7 &  75\\
%NGC~2818      &   139.044 & -36.624 & 4.68 & 9.27 &   -0.059$\pm$0.001\\
%NGC~6134      &   246.953 & -49.161 & 9.36 & 7.29 &    -0.006$\pm$0.010\\
NGC~6253      &   254.778 & -52.712 & 9.70 & 1.69 & 1.58 &  5.36 &  6.88 & -0.18 & 0.141$\pm$0.046 & 5 & 51\\
NGC~6791      &   290.221 &  37.778 & 9.92 & 3.91 & 5.85 & 29.54 &  7.94 &  0.80 & 0.069$\pm$0.008 & 27 & 533 \\
NGC~6819      &   295.327 &  40.190 & 9.36 & 1.76 & 3.91 & 13.32 &  8.03 &  0.41 & 0.143$\pm$0.005 & 10 &  227\\
%NGC~7044      &   318.284 &  42.494 & 9.10 & 2.14 & 2.66 & 18.14 &  8.73 & -0.23 & 0.226$\pm$0.001 & 5& 230 \\
NGC~7142      &   326.290 &  65.782 & 9.55 & 3.10 & 3.92 & 14.79 &  9.25 &  0.40 &-0.020$\pm$0.007 & 8 & 91\\
NGC~7789      &   359.334 &  56.726 & 9.52 & 3.81 & 6.63 & 18.87 &  9.43 & -0.20 & 0.038$\pm$0.004 & 10 & 376\\
%Tombaugh~2    &   105.773 & -20.820 & 9.01 & 1.77 & 3.21 & 24.78 & 15.76 & -1.11 &-0.020$\pm$0.002 & 20& 54 \\
Trumpler~5    &    99.126 &   9.465 & 9.60 & 3.85 & 6.44 & 24.10 & 11.21 &  0.05 & 0.001$\pm$0.016 & 20 & 525\\
%Trumpler~20   &   189.882 & -60.637 & 6.72 & 7.18 &   -0.029$\pm$0.010\\
\hline
\end{tabular}
\end{small}
\end{table*}

\section*{Acknowledgements}
We thank the referee for the thorough reading of the manuscript and
timely suggestions to improve it.

Data for reproducing the figures and analysis in this work will be available upon request
to the author.

%\bibliographystyle{apj}
%\bibliography{paper} % if your bibtex file is called paper.bib

%\input{paper.bbl}

\begin{appendix} 

\section{The $A^+$ index}

In this section, we present color-magnitude diagrams and cumulative distribution
functions of the complete cluster sample. The figures depicts the features
illustrated in Figure~\ref{fig1} (Section~2).

\begin{figure}
\includegraphics[width=\columnwidth]{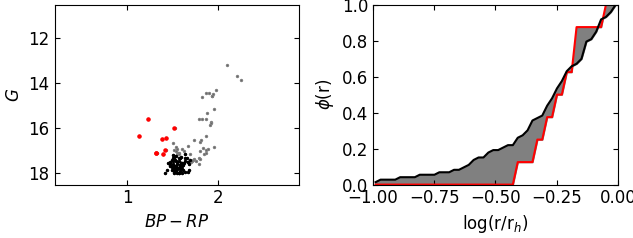}
\caption{Same as Figure~\ref{fig1} for Berkeley~17.}
\label{figa1}
\end{figure}

\begin{figure}
\includegraphics[width=\columnwidth]{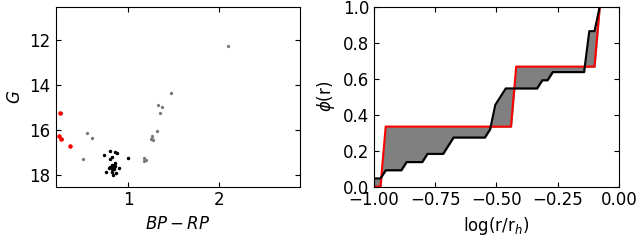}
\caption{Same as Figure~\ref{fig1} for Berkeley~31.}
\label{figa2}
\end{figure}

\begin{figure}
\includegraphics[width=\columnwidth]{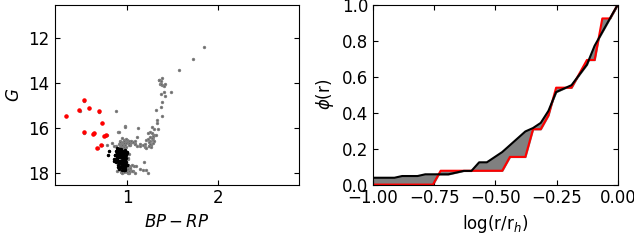}
\caption{Same as Figure~\ref{fig1} for Berkeley~39.}
\label{figa3}
\end{figure}

\begin{figure}
\includegraphics[width=\columnwidth]{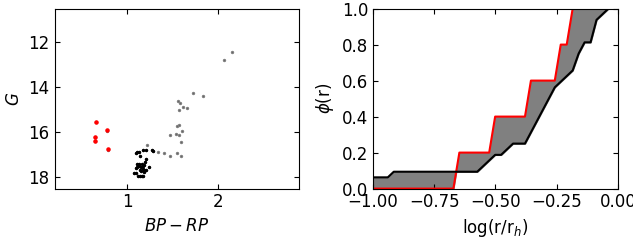}
\caption{Same as Figure~\ref{fig1} for Berkeley~99.}
\label{figa4}
\end{figure}

\begin{figure}
\includegraphics[width=\columnwidth]{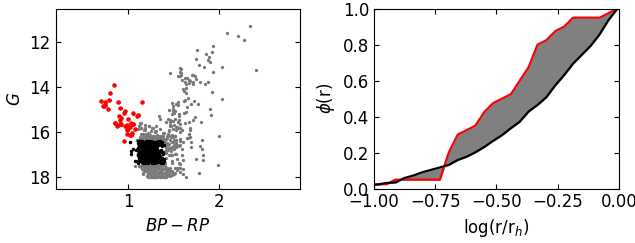}
\caption{Same as Figure~\ref{fig1} for Collinder~261.}
\label{figa5}
\end{figure}

\begin{figure}
\includegraphics[width=\columnwidth]{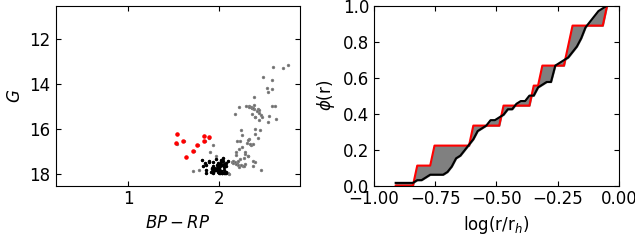}
\caption{Same as Figure~\ref{fig1} for King~11.}
\label{figa6}
\end{figure}

\begin{figure}
\includegraphics[width=\columnwidth]{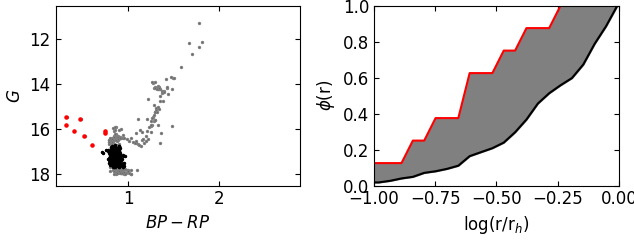}
\caption{Same as Figure~\ref{fig1} for Mellote~66.}
\label{figa7}
\end{figure}

\begin{figure}
\includegraphics[width=\columnwidth]{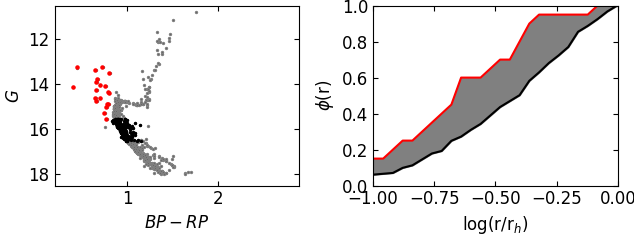}
\caption{Same as Figure~\ref{fig1} for NGC~188.}
\label{figa8}
\end{figure}

\begin{figure}
\includegraphics[width=\columnwidth]{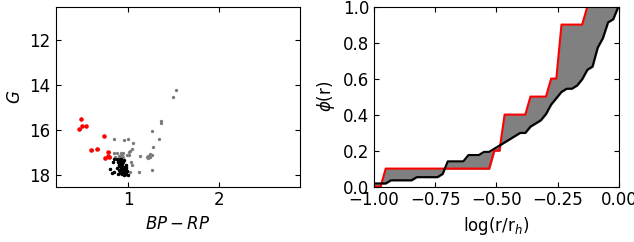}
\caption{Same as Figure~\ref{fig1} for NGC~1193.}
\label{figa9}
\end{figure}

\begin{figure}
\includegraphics[width=\columnwidth]{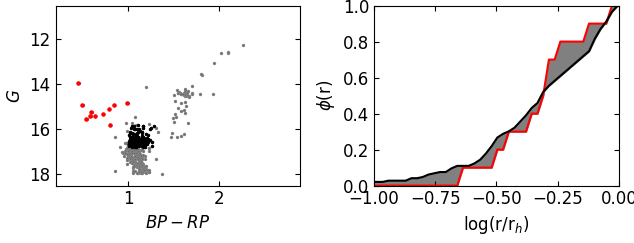}
\caption{Same as Figure~\ref{fig1} for NGC~2141.}
\label{figa10}
\end{figure}

\begin{figure}
\includegraphics[width=\columnwidth]{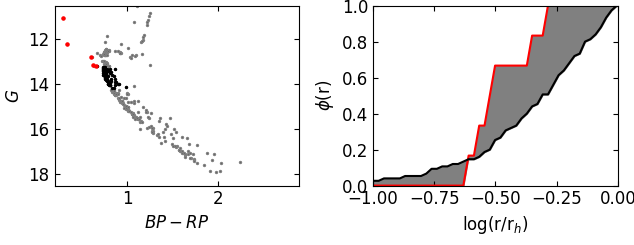}
\caption{Same as Figure~\ref{fig1} for NGC~2682.}
\label{figa11}
\end{figure}

\begin{figure}
\includegraphics[width=\columnwidth]{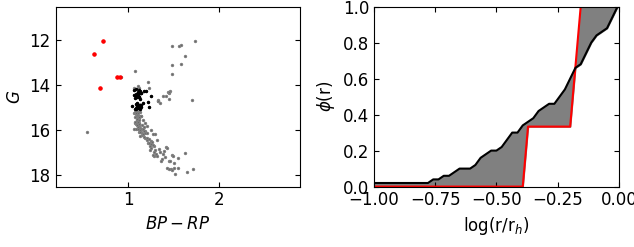}
\caption{Same as Figure~\ref{fig1} for NGC~6253.}
\label{figa12}
\end{figure}

\begin{figure}
\includegraphics[width=\columnwidth]{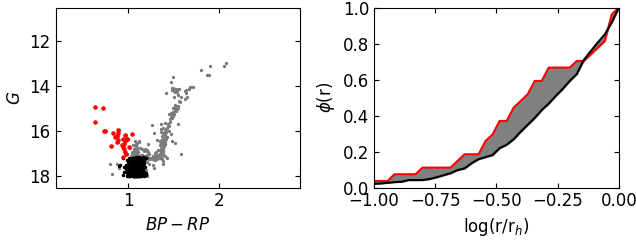}
\caption{Same as Figure~\ref{fig1} for NGC~6791.}
\label{figa13}
\end{figure}

\begin{figure}
\includegraphics[width=\columnwidth]{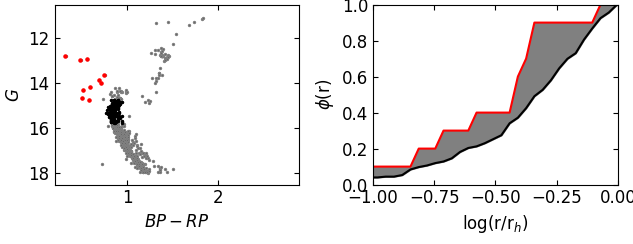}
\caption{Same as Figure~\ref{fig1} for NGC~6819.}
\label{figa14}
\end{figure}

\begin{figure}
\includegraphics[width=\columnwidth]{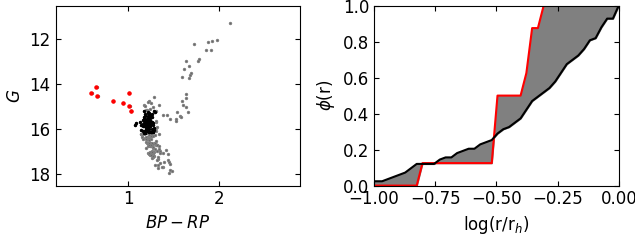}
\caption{Same as Figure~\ref{fig1} for NGC~7142.}
\label{figa15}
\end{figure}

\begin{figure}
\includegraphics[width=\columnwidth]{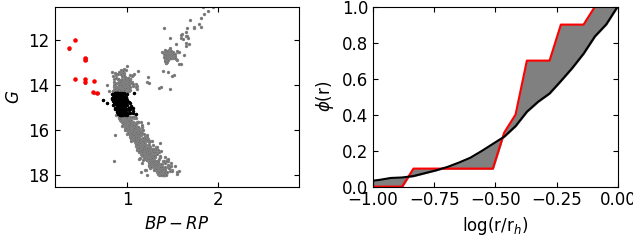}
\caption{Same as Figure~\ref{fig1} for NGC~7789.}
\label{figa16}
\end{figure}

\begin{figure}
\includegraphics[width=\columnwidth]{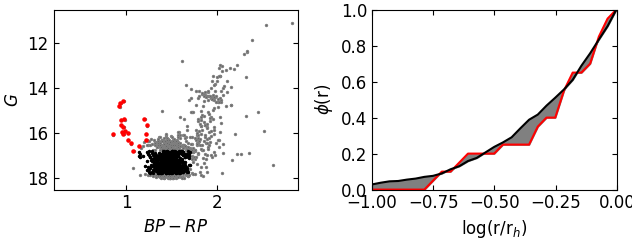}
\caption{Same as Figure~\ref{fig1} for Trumpler~5.}
\label{figa17}
\end{figure}

\begin{figure}
\includegraphics[width=\columnwidth]{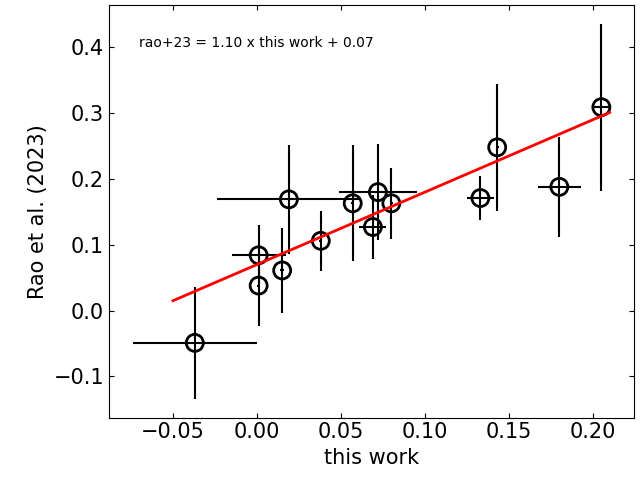}
\caption{Comparison between $A^+$ values derived in this work
and in \citet{raoetal2023a}.}
\label{figa18}
\end{figure}

\end{appendix}

\end{document}